 \documentclass[final,5p,times,twocolumn]{elsarticle}

\usepackage{amssymb}
\usepackage{graphicx}
\usepackage{color,amsmath,amssymb}
\usepackage{color}
\usepackage{comment}
\usepackage{amssymb}
\usepackage{subcaption}
\usepackage{tikz}
\usepackage{longtable}

\newcommand{\C}{CH$^+$-He}
\newcommand{\wn}{cm$^{-1}$}

\journal{JMS}

\begin{document}

\begin{frontmatter}



\title{Rovibrational spectroscopy of the  \C\ and CH$^+$-He$_4$ complexes}


\author[a]{Thomas Salomon}
\author[c]{Jos\'e L. Dom\'enech}
\author[a]{Philipp C. Schmid}
\author[a,b]{Ernest A. Michael} 
\author[a]{Stephan Schlemmer}
\author[a]{Oskar Asvany}
\ead{asvany@ph1.uni-koeln.de}
\address[a]{I. Physikalisches Institut, Universit\"at zu K\"oln, Z\"ulpicher Str.~77,
50937 K\"oln, Germany}
\address[c]{Instituto de Estructura de la Materia  (IEM-CSIC),
Serrano 123, 28006 Madrid, Spain}
\address[b]{Department of Electrical Engineering, University of Chile,  Av. Tupper 2007, Santiago, Chile}


\begin{abstract}
A cryogenic 22-pole ion trap apparatus is used in combination with a table-top pulsed IR source 
to probe weakly bound \C\ and CH$^+$-He$_4$ complexes by predissociation spectroscopy at 4 K. 
The infrared photodissociation spectra of the C–H stretching vibrations are recorded in the 
range of 2720–2800~\wn.  The spectrum of \C\ exhibits perpendicular transitions of 
a near prolate top with a band origin at 2745.9~\wn, and thus 
confirms it to have a T-shaped structure. For CH$^+$-He$_4$, the  C-H stretch along the symmetry axis of this oblate top results in  parallel transitions. 
\end{abstract}


\end{frontmatter}


\section{Introduction \label{intro}}

Molecular complexes consisting of a cation and a weakly bound neutral partner have been investigated 
by action spectroscopy in the last four decades. 
These techniques were  primarily invented and used to obtain spectroscopic information about the 
bare cation itself~\cite{boo95,bie00,bru03,jas13,jas15}, but
more recently the focus shifted also towards the weak interaction between the constituents
and the corresponding shallow potential energy surface (PES), in particular for floppy cation-helium complexes.
For instance,  IR spectra, and, more recently,  also partly rotational spectra of 
N$_2$H$^+-$He~\cite{niz95b}, HCO$^+-$He~\cite{niz95,sal19}, OH$^+-$He~\cite{rot98}, CH$_3^+-$He~\cite{olk99,toe18}, 
N$_2^+-$He~\cite{bie92}, {\color{red} NH$_4^+-$He~\cite{lak01},
O$_2$H$^+-$He~\cite{niz97,koh18},}
H$_3^+-$He~\cite{sav15} and H$^+-$He$_n$~\cite{asv19,toe20} have been explored.

\begin{figure}[t!]
\begin{center}
 \includegraphics[width=.4\textwidth]{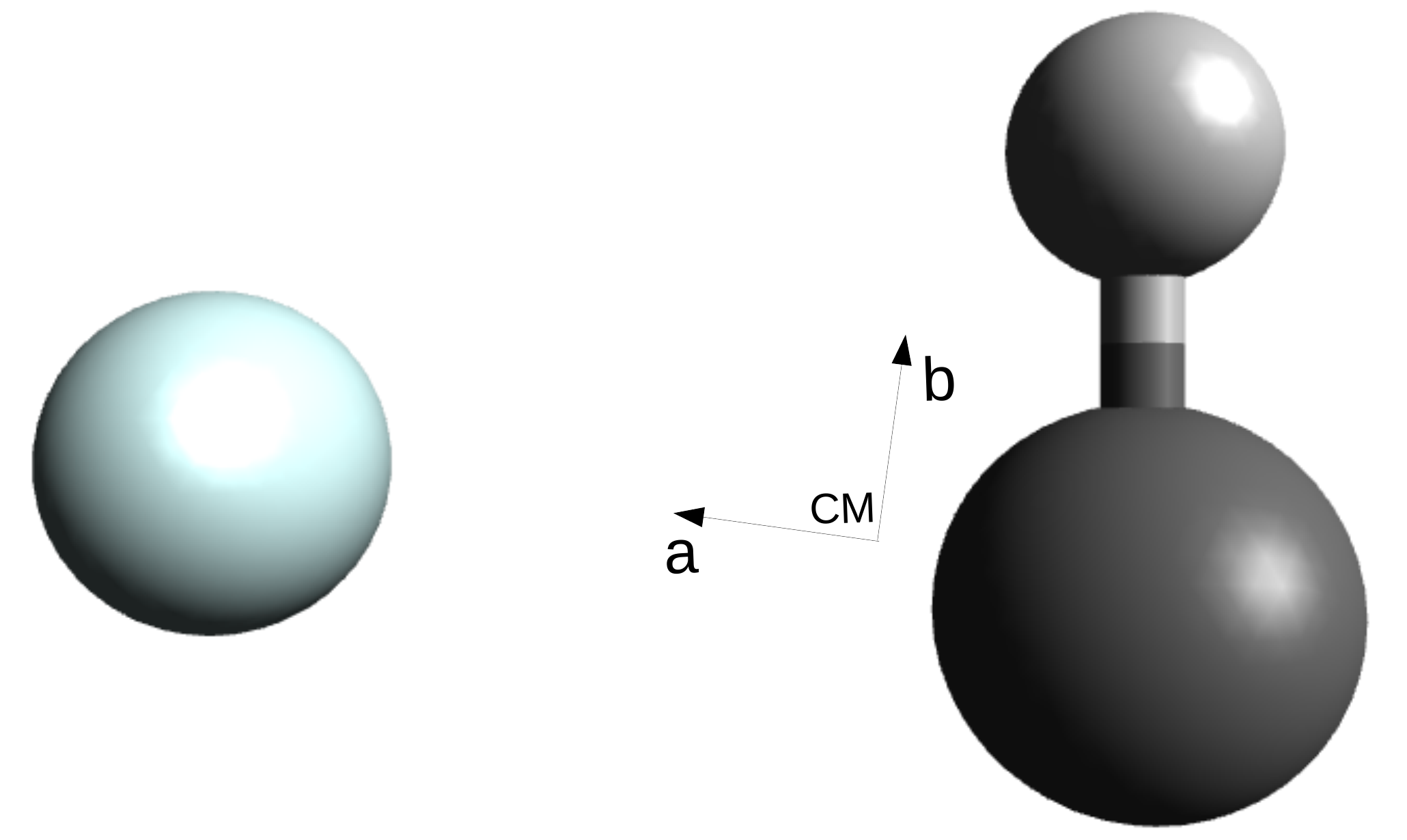}
 \end{center}
 \caption{\label{sketch}   
Sketch of \C\ depicting its center of mass and its principal axes of inertia. 
}
\end{figure}

CH$^+$  was the very first molecular ion known to exist in interstellar space,
detected by its electronic transitions~\cite{dun37,dou41}, and since then it has also been detected
by its rotational~\cite{cer97,fal05,men11,nag13}, 
and vibrational~\cite{neu20} transitions (the latter only very recently). 
In the  astronomically important collision between CH$^+$  
and He  the weakly bound \C\ complex is formed as an intermediate.
{\color{red} Up to date, the  \C\ complex has been investigated only theoretically,
first by Hughes and  Nagy-Felsobuki~\cite{hug97},  and later
 by Meuwly and Wright~\cite{meu00}. }
These authors predict a T-shape structure rather than a linear one, see Fig.~\ref{sketch}.
Detailed knowledge of the CH$^+$  - He PES is  important to be able to predict
inelastic collision rates in space~\cite{sto08,ham08}. 
Determination of its bound states by rotational or rovibrational spectroscopy, together with theoretical work,
is a very exact way of exploring the underlying PES.

In a previous work aimed at high-resolution detection of the C-H stretching fundamental of the bare
CH$^+$ cation~\cite{dom18}, we serendipitously found seven Lorentzian lines which we ascribed to \C.
In this work, we complete the search for the lines of the complex using the same ion trap machine, and
apply a lower-resolution pulsed laser to record the  infrared photodissociation (IRPD) spectrum of \C. In addition, we also 
measure the IRPD spectrum of the symmetric complex CH$^+$-He$_4$.

\section{Experimental details}
\label{sec:expt}

\begin{figure}[t!]
 \includegraphics[width=.48\textwidth]{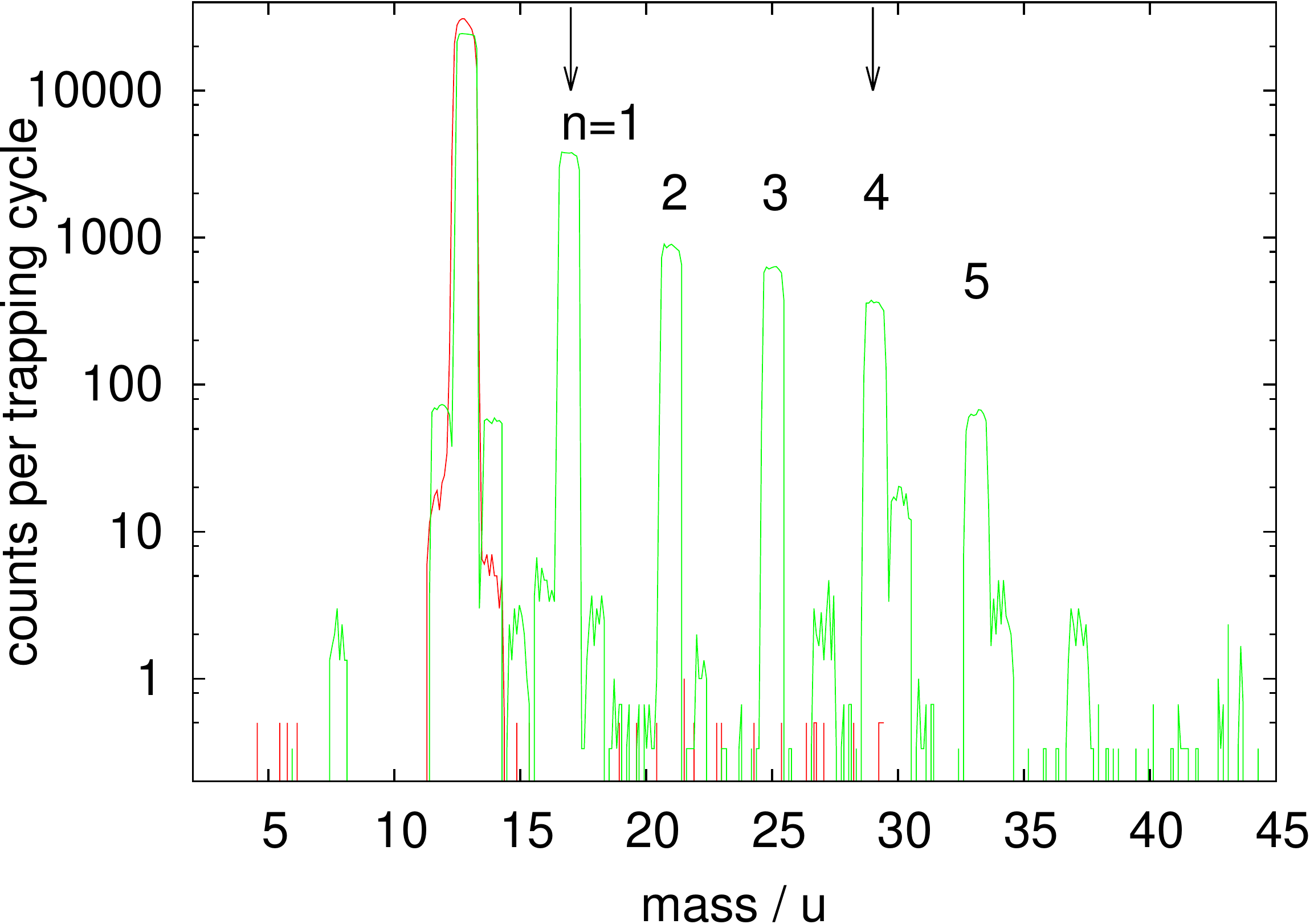}
 \caption{\label{mass_spec}   
 Mass spectra showing the trap content at time of injection (red trace) 
 and the formation of CH$^+$-He$_n$ by  trapping for 700~ms 
 in a constant 
 high-density He environment (green trace). The ion on mass 30~u is most probably due to the parasitic reaction CH$^+$ + H$_2$O $\rightarrow$
 CH$_2$O$^+$ + H. A similar spectrum for CD$^+$-He$_n$ has been given in Reference \cite{bru17}. 
 The arrows indicate the mass channels ($n=1,4$) in which the IRPD-detection in this work occurs.
}
\end{figure}

The IRPD experiments of \C\ and CH$^+$-He$_4$ are carried out in the ion trapping machine COLTRAP~\cite{asv14}.
A 1:1 mixture of helium and methane (Linde 5.5) is used as a precursor to 
create CH$^+$ ions by electron impact ionization in an external storage ion source (electron energy about 30~eV). Helium is used in the precursor mixture because  
ionized helium is known to promote the fragment ion production~\cite{tho19}. 
Subsequent mass-filtering ($m = 13$~u) in a linear quadrupole ensures that 
a pure sample of several ten thousand CH$^+$ ions enters the 4~K 22-pole ion trap~\cite{asv10}.
About 40~ms before the ion bunch
reaches the trap, He buffer gas is injected into the trap
via a piezo valve and is thus thermalized by collisions with the trap walls.
This leads to the cooling of the incoming ions by collisions with the buffer gas. 
The formation of \C\ and higher complexes CH$^+$-He$_n$ via ternary collision 
processes is promoted by the low trap temperature and the high instantaneous He number density.
In the mass spectrum of Fig.~\ref{mass_spec} one  can see that the complexes with $n=1-4$ are readily formed,
whereas complexes with $n=5$ or even $n=6$ have  less abundance, 
 in agreement with a shell closure at a four-membered He-ring around the CH$^+$ molecular axis, 
 as suggested earlier~\cite{bru17}.  

For IRPD spectroscopy, the shown ion ensemble is trapped for a  
period  of 2.5~s and irradiated by the pulsed infrared radiation,
leading to excitation upon resonance and subsequent 
destruction of the  complexes 
{\color{red}(\C\ has a dissociation energy of $D_0=243$~\wn~\cite{meu00}).}
After extracting all ions from the trap, the ions are mass-filtered in the \C\ or 
CH$^+$-He$_4$ mass channels 
(\mbox{$m = 17$ or 29~u}, see arrows in Fig.~\ref{mass_spec})
by a second linear quadrupole and efficiently counted by a Daly-type detector.
The predissociation spectrum is recorded by counting the number of these complexes 
(typical counts are on the order of 5000) as a function of the IR excitation frequency.
In that way, the resonant absorption of IR photons is observed as a dip in the  counts, 
see lower part of Fig.~\ref{spec}. As long-term fluctuations of the complex counts were not severe, we omitted to 
use a laser shutter for normalization purposes.
%

The pulsed IR radiation is produced by a table-top
LaserVision  optical parametric oscillator/amplifier (OPO/OPA) system. 
The OPO/OPA system is pumped with a 1064 nm Nd:YAG laser (Continuum Surelite-Ex) operating at 10 Hz
and with maximum pulse energies of 600~mJ. 
This system has a bandwidth of  0.1~\wn\ when seeded.
The IR laser wavelength is monitored with a wavemeter (HighFinesse WS-5) with a 
manufacturer-stated accuracy of 0.1~\wn. 
The IR beam passes through the trapping machine via two Brewster window assemblies~\cite{asv20b} 
adjusted for vertical polarization and containing BaF$_2$ windows.
Typical pulse energies are on the order of 3~mJ (in seeded mode). 

\section{Rovibrational spectroscopy of \C}
\label{spectr1}

During our former work~\cite{dom18} we serendipitously recorded seven 
\C\ lines and even a single $^{13}$CH$^+$-He line. The latter, at  2758.548404(5)~\wn, 
is depicted in Fig.~1 of  reference~\cite{dom18}. 
The seven unpublished and initially unassigned lines of \C\ are listed in Table~\ref{list}.
The use of a high-resolution narrow-band cw IR source allowed to determine their positions 
and Voigt line profiles with high accuracy and precision 
(the accuracy of the used wavemeter was 30~MHz, the precision even better). 
Thus the lifetimes of the vibrationally excited states, in the order of 1-2~ns, could be determined (see Table~\ref{list}). 
Two  lines were recorded at even higher precision and accuracy (about 80~kHz) using a frequency comb system~\cite{asv12}.

Similar to CH$^+$ ($^1 \Sigma$), the  complex \C\ is a closed-shell molecule with a singlet
electronic ground state. {\it Ab initio} calculations~\cite{hug97,meu00} predicted 
its equilibrium structure to be non-linear.  
It has rather a T-shaped equilibrium structure with the helium atom in a distance 
of $R_e=2.28$~\AA\ and angle of $\theta_e=83^\circ$ relative to the C-H center of mass and axis, respectively (see Fig.~\ref{sketch}).
Its symmetry group is thus C$_s$ and its ground electronic state is designated $^1A'$. 
Its calculated rotational constants~\cite{meu00} reveal that it is a near prolate top.
As the C-H stretch investigated here points almost perpendicular to the symmetry axis of 
the prolate top (the symmetry axis is the a-axis, the C-H transition moment points almost along the b-axis), a perpendicular transition of a near prolate top can be expected.

\begin{figure}[t!]
 \includegraphics[width=.48\textwidth]{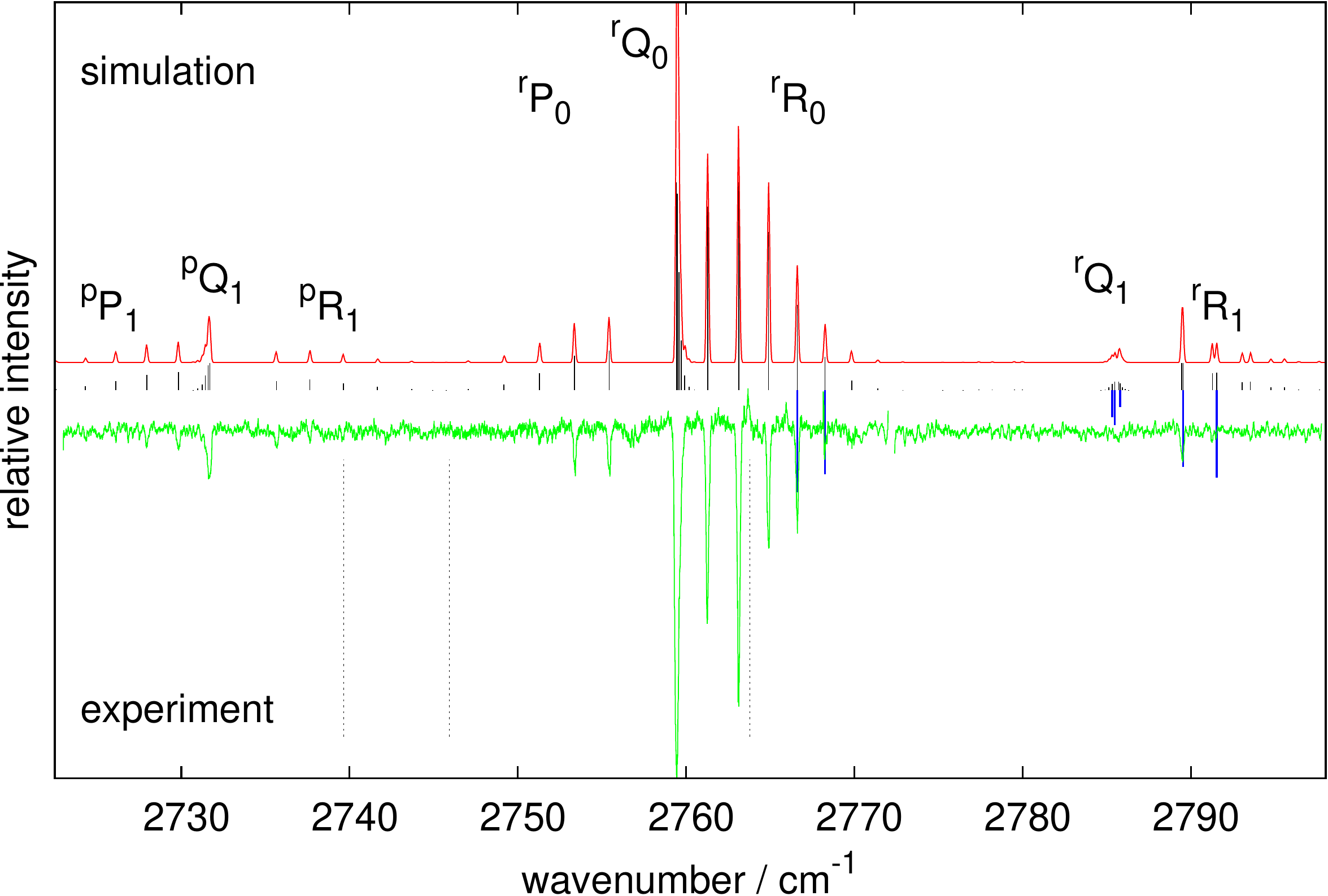}
 \caption{\label{spec}   
Rovibrational spectrum of the C-H stretching vibration of \C. upper part: PGOPHER simulation~\cite{wes17} displaying stick spectrum (black) and convoluted spectrum (red). 
lower part: experimental data of this work (green), 
including lines from previous work as given in Table 1 (blue sticks). 
{\color{red}The band origin of CH$^+$ at 2739.670097(5)~\wn~\cite{dom18}, as well as the blue-shifted 
band origins of \C\ and CH$^+$-He$_4$ (see Fig.~\ref{spectr2})
are indicted by dashed vertical lines.}
Branches of \C\ are designated by the notation $^{\Delta K} \Delta J _{K}$.
}
\end{figure}

The spectrum obtained in this work indeed resembles that of a
perpendicular transition of a near prolate top.
It is shown in the lower part of Fig.~\ref{spec},
together with the seven high-resolution lines obtained in our earlier work, depicted as blue
sticks. As at least two lines of the high-resolution data set could be detected and well separated  in this work, 
they have been used to re-calibrate the spectrum of this work (it has been shifted up by 0.13~\wn, 
which is well within the accuracy of the used wavemeter). The experimental spectra  are accompanied by 
PGOPHER~\cite{wes17} simulations in the upper part of Fig.~\ref{spec}, one with narrow sticks and one convoluted with 
a Gaussian laser bandwidth of 0.1~\wn\  representative of the current experimental conditions.
The fit of the simulation to the spectroscopic data yields a rotational temperature of about 10~K
and spectroscopic constants as summarized in Table~\ref{param}. 
Fortunately, we were able to resolve one asymmetry splitting in the $^rR_1$ branch (though with low signal-to-noise ratio),
which in turn enabled us to assign the formerly measured high-resolution lines with confidence,
so that  the parameter $B-C$ could be determined.  
All assigned lines are collected in Table~\ref{list}.
The weighted RMS of the fit is 0.29, which shows that our error limits given for the lines 
of this work are very conservative.

{\color{red}
For the determination of the structure of \C\ we assume the C-H backbone to be unchanged w.r.t.\  CH$^+$.
From the ground state rotational constant $B_{CH} = 13.9313576(1)$~\wn~\cite{ama10b,dom18} of  CH$^+$ we obtain a 
C-H bond length of $r_0=1.141$~\AA. As the principal a-axis of \C\ passes trough the He atom and the center of mass of CH (see Fig.~\ref{sketch}),
the corresponding rotational constant $A_0=15.15$~\wn\ is determined by the tilt of the CH-unit towards this axis,
i.e. by the Jacobi angle, for which we calculate $\theta_0 = 73.52^\circ$ for the ground vibrational state.
Using also the rotational constant $B_0=1.0184$~\wn\ (see Table~\ref{param}) we further estimate  $R_0=2.319$~\AA\ 
(the distance between He and the center of mass of the CH subunit), in good agreement 
with the average value $R_0=2.4$~\AA\ given by Meuwly and Wright~\cite{meu00} for the ground vibrational state. In a similar way, using $r_1=1.162$~\AA\ 
for the vibrationally excited state probed in this work, we obtain an angle $\theta_1 = 74.79^\circ$ and a distance which is only slightly larger, $R_1=2.321$~\AA.
}

\begin{table*}[t] 
\begin{center}
 \caption{\label{list} Rovibrational transitions of the $\nu_{1}$ band of \C.
Experimental uncertainties are given in parentheses. }
 \begin{tabular}{lclr@{}lr@{}l}
\hline
            & $J'_{Ka'Kc'}  \leftarrow J"_{Ka"Kc"}$  &  $\tilde{\nu}$ / \wn   & \multicolumn{2}{c}{Obs.-Calc.} & \multicolumn{2}{c}{$\tau$ / ns $^a$}  \\
\hline
$^pP_1(3)$  & $2_{02} \leftarrow 3_{13}$   & 2726.14(5)         &  0&.015     &   \\
$^pP_1(2)$  & $1_{01} \leftarrow 2_{11}$   & 2727.97(5)         & -0&.003 &  \\
$^pP_1(1)$  & $0_{00} \leftarrow 1_{11}$   & 2729.86(5)         &  0&.002 &\\
$^pQ_1$     &                              & 2731.7                        \\
$^pR_1(1)$  & $2_{02} \leftarrow 1_{11}$   & 2735.69(5)         &  0&.012 &\\
$^pR_1(2)$  & $3_{03} \leftarrow 2_{12}$   & 2737.67(5)         &  0&.010 &\\
$^pR_1(3)$  & $4_{04} \leftarrow 3_{13}$   & 2739.63(5)         & -0&.031 &\\
\hline
$^rP_0(4)$  & $3_{13} \leftarrow 4_{04}$   & 2751.32(5)         &  0&.027      &     \\
$^rP_0(3)$  & $2_{12} \leftarrow 3_{03}$   & 2753.40(5)         &  0&.015  & \\
$^rP_0(2)$  & $1_{11} \leftarrow 2_{02}$   & 2755.46(5)         &  0&.019 &\\
$^rQ_0$     &                              & 2759.45                 \\
$^rR_0(0)$  & $1_{11} \leftarrow 0_{00}$   & 2761.27(5)         & -0&.034 &\\
$^rR_0(1)$  & $2_{12} \leftarrow 1_{01}$   & 2763.13(5)         & -0&.005 &\\
$^rR_0(2)$  & $3_{13} \leftarrow 2_{02}$   & 2764.92(5)         &  0&.018  &\\
$^rR_0(3)$  & $4_{14} \leftarrow 3_{03}$   & 2766.607175(8) $^{b,c}$&  0&.0     &  1&.50(2) \\
$^rR_0(4)$  & $5_{15} \leftarrow 4_{04}$   & 2768.254(1) 	$^b$ &     0&.0     &  1&.63(7) \\
\hline
$^rQ_1(3)$  & $3_{21} \leftarrow 3_{12}$   & 2785.315(1) 	$^b$ & 	   0&.0     &	1&.4(2)  \\
$^rQ_1(2)$  & $2_{20} \leftarrow 2_{11}$   & 2785.468(1) 	$^b$ & 	   0&.0     &	1&.5(1)  \\
$^rQ_1(3)$  & $3_{22} \leftarrow 3_{13}$   & 2785.791(1) 	$^b$ & 	   0&.0     &	1&.8(5)  \\
$^rR_1(1)$  & $2_{20} \leftarrow 1_{11}$   & 2789.534(1) 	$^b$ & 	   0&.0     &	1&.61(6) \\
$^rR_1(2)$  & $3_{21} \leftarrow 2_{11}$   & 2791.51941(2)  $^{b,c}$ & 0&.0     &	1&.65(3) \\
\hline
 \end{tabular}
 \end{center}
 $^a$ The lifetimes are determined by extracting the Lorentzian width in a Voigt fit, assuming the 
Gaussian contribution to correspond to a temperature of 14~K\\
$^b$ The seven \C\ lines with higher precision have been detected in our previous work~\cite{dom18}. \\
$^c$ lines with very high precision and accuracy have been measured with a frequency comb system. \\
\end{table*}

\begin{table}[t]
\begin{center}
 \caption{\label{param} Spectroscopic parameters for  \C\ and CH$^+$-He$_4$. 
All values in cm$^{-1}$. Experimental uncertainties are given in parentheses.
Spectroscopic fits were done with the program PGOPHER~\cite{wes17}. }
\begin{tabular}{lr@{}llr@{}lr@{}l}
     & \multicolumn{3}{c}{ab initio $^a$}          &  \multicolumn{4}{c}{this work}   \\
     & \multicolumn{2}{c}{ground}  &  $v_1=1$      & \multicolumn{2}{c}{ground}  & $v_1=1$          \\
\hline
\multicolumn{3}{l}{\C, C$_s$}  \\
$\nu$ &     & & 2869 $^b$                          &      &           &  2745 &.94(1)       \\[0.25cm]
$A$       & 15. & 28  &                            &   15 &.15(1)     & 14 &.435(2)           \\
$\bar{B}$ &  0. & 93  &                            &   0  &.9785(2)   & 0 &.9708(4)            \\
$B-C$     &  0. & 074 &                            &   0 & .07972(9)  & 0 &.091(3)             \\[0.15cm]
$D_J$ $\times 10^3$     & \multicolumn{3}{c}{}     &  0  & .23(1)    & 0 & .13(2)         \\
\hline
\hline
\multicolumn{3}{l}{CH$^+$-He$_4$, C$_{4v}$}  \\
$\nu$ &     & &                                    &      &    &  2763 &.77(5)       \\[0.25cm]
$B$   &     &  &                                   &   0  &.378(1)   & 0 &.375(1)            \\
\hline
\end{tabular}
\end{center}
$^a$ Reference~\cite{meu00} \\
$^b$ harmonic value
\end{table}

\section{Rovibrational spectroscopy of CH$^+$-He$_4$}
\label{spectr2}

In order to get more insight into the structure of the helium complexes, 
we also investigated CH$^+$-He$_4$ by IRPD. We chose this complex instead of 
CH$^+$-He$_2$ or CH$^+$-He$_3$,  as the latter are suspected to be floppy with a complicated 
spectrum (and we postpone their investigation to a later work). 
CH$^+$-He$_4$, on the other hand, is predicted to be an oblate symmetric top~\cite{sol16}, with a 
4-membered ring of helium atoms surrounding the C-H axis. 
It thus belongs to the symmetry group C$_{4v}$.
The IRPD spectrum of 
CH$^+$-He$_4$, measured by observing the mass channel $m = 29$~u (see Fig.~\ref{mass_spec}) 
as a function of laser wavenumber, is shown in Fig.~\ref{spectr2}. It exhibits 
parallel transitions ($\Delta K = 0$) of a symmetric top molecule, thus confirming the proposed structure.
At the resolution of this experiment, only the rotational constant $B$ (the one perpendicular to the symmetry 
axis of the top) and the band origin can be determined with confidence, and are included in Table~\ref{param}. 
{\color{red} As expected, the band origin, at  2763.77(5)~\wn, 
is further blue shifted w.r.t.\ the ones of CH$^+$ and \C\ 
(see also the dashed vertical lines in Fig.~\ref{spec}).
As anticipated by theoretical work~\cite{sol16}, these blue shifts seem very linear, 
and we obtain a value of about 6~\wn\ per additional 
He atom. Therefore, we may predict band origins of  2751.7~\wn\ and 2757.7~\wn\
for the floppy CH$^+$-He$_2$ and CH$^+$-He$_3$, respectively.
}

The experimental rotational constant $B=0.378$~\wn\ is in good agreement with the proposed structure of 
CH$^+$-He$_4$. Assuming this molecule to be formed by filling up the 4-membered ring with He atoms
and having all a distance  of about 2.22~\AA\ to the CH-axis as calculated for \C,
we obtain a rotational constant $B=0.384$~\wn. 
This value even suggests that the distance of the He 
atoms to the CH-axis is slightly larger for CH$^+$-He$_4$ than for \C.
{\color{red} In summary, this is consistent with the picture that in the CH$^+$-He$_n$ series, every additional He atom strengthens the C-H bond, leading 
to the mentioned additive blue shift of the band origin
of about 6~\wn/He atom,  while the CH$^+$ - He bonds get weaker, leading 
to slightly larger distances of the He atoms to the CH-axis.}

\begin{figure}[t!]
 \includegraphics[width=.48\textwidth]{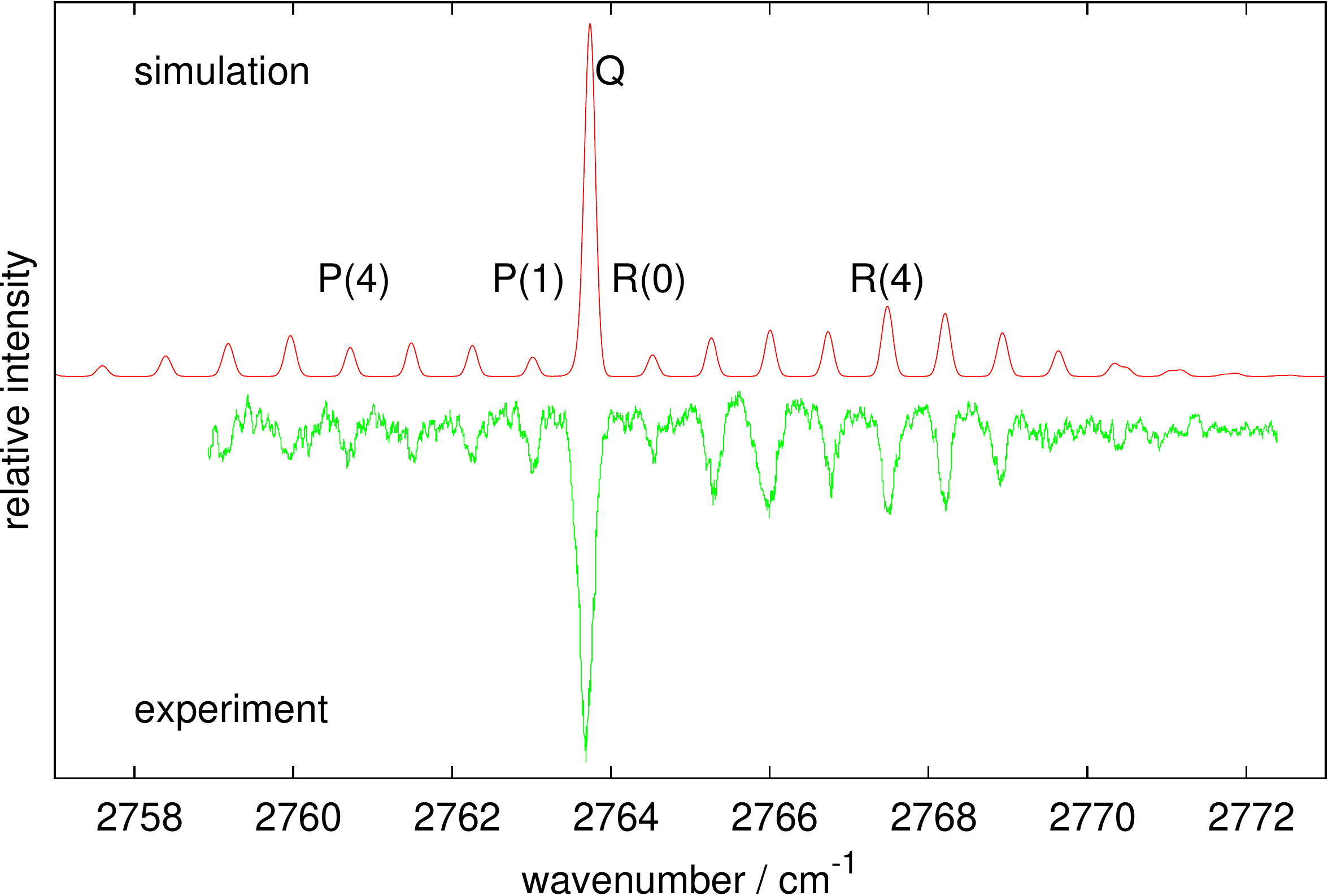}
 \caption{\label{spec2}   
Rovibrational spectrum of the C-H stretching vibration of CH$^+$-He$_4$ 
(upper panel: PGOPHER simulation~\cite{wes17} in red, 
lower panel: experimental data in green). 
Some crosstalk of this spectrum into that of  \C\ can be observed in Fig.~\ref{spec}
{ \color{red}
(see e.g. the rightmost dashed vertical line in Fig.~\ref{spec},  
which is the position of the Q-branch of CH$^+$-He$_4$). }
In the simulation, we took into account that only the nuclear spin weights of the 
irreducible representations A$_1$ and A$_2$ are non-zero. 
}
\end{figure}


\section{Conclusions and Outlook}
\label{sec:conclusion}

Using a combination of a cryogenic ion trap machine with a 
tunable table-top pulsed IR source
the $\nu_1$ C-H stretches  of the cation-helium complexes \C\ and CH$^+$-He$_4$
have been recorded in the 3~$\mu$m region. 
{\color{red} To the best of our knowledge,  CH$^+$-He$_4$ is the largest cation-helium complex 
($n=4$) for which a rotationally resolved IR spectrum has been reported.}
The investigation confirms the proposed structures of \C~\cite{meu00} and CH$^+$-He$_4$~\cite{sol16}, 
with the helium atoms attached 
to the C-H backbone in a T-shaped manner,  by multiple evidence:
i) the mass spectrum shows a magic number at 4 helium atoms and thus supports a 4-membered ring around the CH$^+$ axis, 
{\color{red}
ii) the relatively long lifetime of the vibrationally excited \C\ complex of about 1.6~ns
is consistent with (but does not necessarily prove) the fact that the He-bond does not point along the C-H axis,
iii)  the band origins  of the C-H stretch vibrations of \C\ and CH$^+$-He$_4$, when compared to 
the one of CH$^+$ (2739.670097(5)~\wn~\cite{dom18}), 
exhibit blue shifts which are quite small and show a regular additive behavior},
and finally, 
iv) the full spectral analyses given in the preceding sections reveals  \C\ and  CH$^+$-He$_4$
to have rotational constants which are in close agreement with the 
proposed structures.
It is astonishing that such potentially floppy molecules with weak bonds of the helium atoms to the CH$^+$ core
present themselves as  texbook-like  rigid near-prolate  and oblate symmetric tops, respectively, at least at 
the resolution of the current experiment. 
Further high-resolution experiments are necessary to reveal deviations from a perfect
rigid rotor model and to explore the shallow potential energy surface via the 
determination of the higher order centrifugal distortion constants.

{\color{red}
The \C\ complex can be compared to other complexes which have a linear cation core.
Typically, complexes of the type AH$^+$-He (with AH$^+$ being the linear cation core)
exhibit a linear proton-bound structure, leading to 
a redshift of the A-H vibrational frequency upon complexation with He, as witnessed for the mentioned   systems
HCO$^+-$He~\cite{niz95,sal19}, OH$^+-$He~\cite{rot98}, N$_2$H$^+-$He~\cite{niz95b},
HeH$^+-$He~\cite{toe20},
and H$_2^+-$He~\cite{kon19}. To the best of our knowledge,
\C\ is the only such complex with a T-shaped structure.
\C\ and CH$^+$-He$_4$ are thus somewhat more similar to the 
H$^+-$He$_n$ series of complexes, for which 
the symmetric linear complex with $n=2$,  He-H$^+$-He, has been determined as the 
central core, and with $n=3$ having a T-shaped structure 
and $n=6$ a 4-membered helium ring around this  core.
While low-resolution IR studies could confirm these motifs~\cite{asv19},
high-resolution studies for $n \ge 3$ are currently hampered by lifetime 
broadening~\cite{toe20}.}

\begin{figure}
 \includegraphics[width=.48\textwidth]{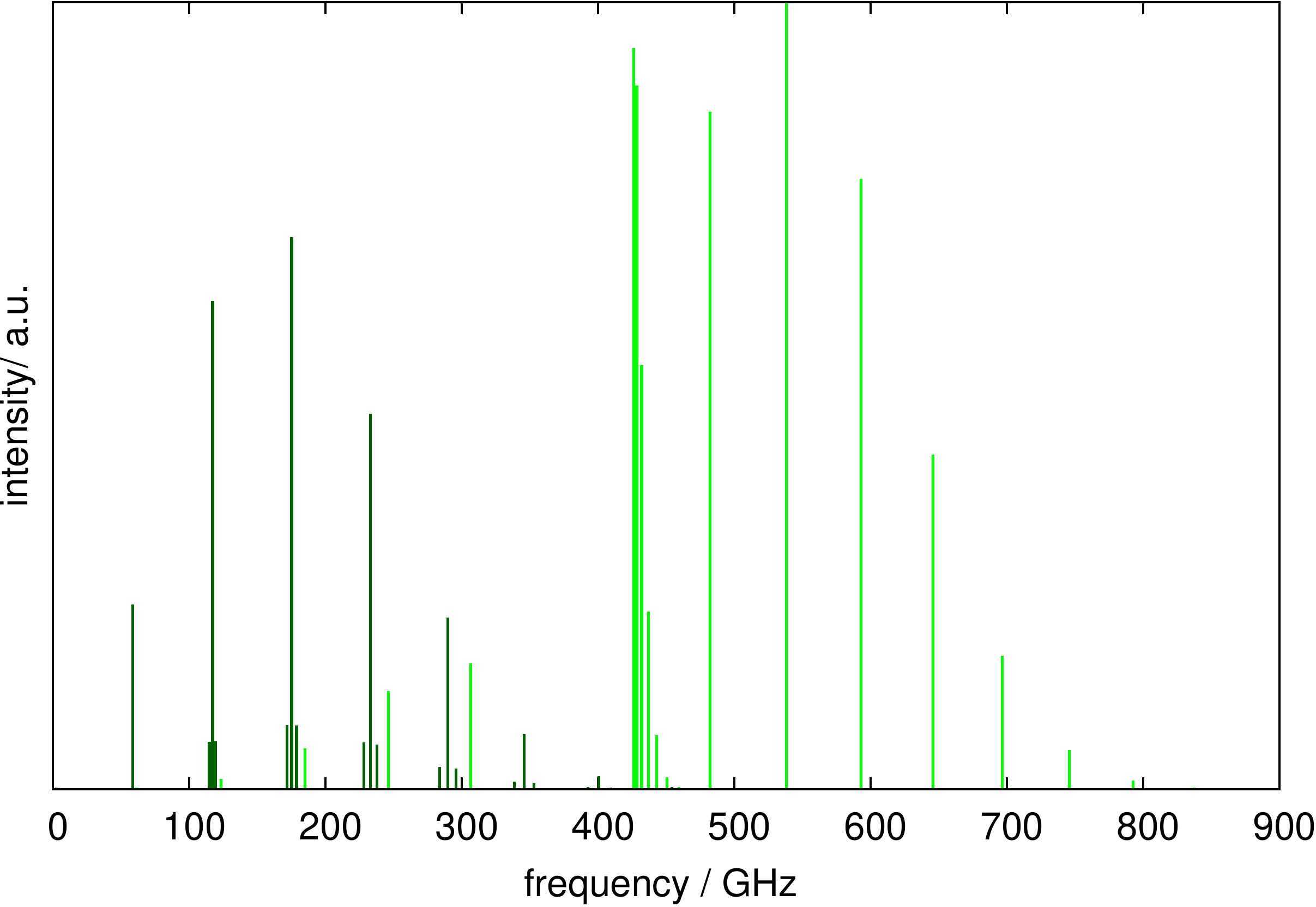}
 \caption{\label{rotpredict} Prediction of the expected rotational spectrum of \C\ at $T=10$~K, 
based on the spectroscopic parameters given in Table~\ref{param}. 
The a-type  ($\Delta K_a = 0$, $\Delta K_c = \pm 1$) and the 
b-type  ($\Delta K_a = \pm 1$, $\Delta K_c = \pm 1$) transitions are shown in dark green and light green, respectively.
}
\end{figure}

More detailed information about the structure of \C\ can also
be obtained by the investigation of its ground vibrational state. 
Rotational spectroscopy of \C\ is feasible as it possess a permanent dipole 
moments along both the a- and b-axes (see Fig.~\ref{sketch}). 
We expect the dipole moment along the a-direction to be particularly strong, 
as the charge sits mainly on the CH$^+$ moiety which is separated 
by more than 2~\AA\  from the He atom.  
A rough prediction of the a-type and b-type rotational  spectra can be found in Fig.~\ref{rotpredict}.
High-resolution rotational spectroscopy of \C\ can be achieved by combining 
the IRPD spectroscopy presented in this work with a rotational photon in a 
double-resonance scheme, as recently demonstrated for the complexes CH$_3^+$-He and 
HCO$^+$-He~\cite{toe18,sal19}. 
As shown in those works, the rotational spectroscopy of  loosely bound complexes is not 
limited by lifetime and broadening issues, so that accurate quantum state  information on the kHz-level 
can be obtained.

\section*{Acknowledgements}
This work has been supported via
Collaborative Research Centre 956, sub-project B2, funded by the Deutsche Forschungsgemeinschaft (DFG, project ID 184018867)
as well as  DFG SCHL 341/15-1 (``Cologne Center for Terahertz Spectroscopy'') and AS 319/2-2. 
JLD acknowledges partial financial support from the Agencia Estatal de Investigación (AEI) through grant FIS2016-77726-C3-1-P
and from the European Research Council through grant agreement ERC-2013-SyG-610256-NANOCOSMOS.
We thank the anonymous reviewer for a thorough check of the submitted manuscript.


\bibliographystyle{elsarticle-num} 

\end{document}